\documentclass[letterpaper, 10 pt, conference]{ieeeconf}  
\usepackage{amsmath}
\usepackage{amsfonts}
\usepackage{mathtools}
\usepackage{bbold}
\usepackage{graphicx}
\usepackage{color}
\usepackage{tikz}
\usetikzlibrary{positioning}
\usepackage{tikzscale}
\usepackage{import}
\usetikzlibrary{backgrounds,calc,shadings,shapes.arrows,arrows,shapes.symbols,shadows,positioning,decorations.markings,backgrounds,arrows.meta}
\usepackage{caption}
\usepackage{stackengine}
\usepackage{hyperref}

\usepackage{array}
\usepackage{booktabs}
\usepackage{multirow}
\usepackage{siunitx}
\usepackage{arydshln}
\makeatletter
\def\adl@drawiv#1#2#3{%
        \hskip.5\tabcolsep
        \xleaders#3{#2.5\@tempdimb #1{1}#2.5\@tempdimb}%
                #2\z@ plus1fil minus1fil\relax
        \hskip.5\tabcolsep}
\newcommand{\cdashlinelr}[1]{%
  \noalign{\vskip\aboverulesep
           \global\let\@dashdrawstore\adl@draw
           \global\let\adl@draw\adl@drawiv}
  \cdashline{#1}
  \noalign{\global\let\adl@draw\@dashdrawstore
           \vskip\belowrulesep}}
\newcommand{\cdashlinelrdotted}[1]{%
  \noalign{\vskip\aboverulesep
           \global\let\@dashdrawstore\adl@draw
           \global\let\adl@draw\adl@drawiv}
  \cdashline{#1}[.4pt/1pt]
  \noalign{\global\let\adl@draw\@dashdrawstore
           \vskip\belowrulesep}}
\makeatother

\definecolor{darkblue}{HTML}{1f4e79}
\definecolor{lightblue}{HTML}{00b0f0}
\definecolor{salmon}{HTML}{ff9c6b}
\definecolor{dodgerblue}{rgb}{0.12, 0.56, 1.0}
\definecolor{frenchblue}{rgb}{0.0, 0.45, 0.73}
\definecolor{green(pigment)}{rgb}{0.0, 0.65, 0.31}
\definecolor{macaroniandcheese}{rgb}{1.0, 0.74, 0.53}
\definecolor{arylideyellow}{rgb}{0.91, 0.84, 0.42}
\definecolor{pansypurple}{rgb}{0.47, 0.09, 0.29}
\definecolor{glaucous}{rgb}{0.38, 0.51, 0.71}
\definecolor{hanblue}{rgb}{0.27, 0.42, 0.81}
\definecolor{newblue}{rgb}{0.56, 0.67, 0.85}
\definecolor{newgreen}{rgb}{0.67, 0.82, 0.57}
\definecolor{fireenginered}{rgb}{0.81, 0.09, 0.13}

\IEEEoverridecommandlockouts                              

\title{\LARGE \bf
EEG-GMACN: Interpretable EEG Graph Mutual \\ Attention Convolutional Network
}

\author{
Haili Ye$^{1}$, 
Stephan Goerttler$^{1,2}$, 
Fei He$^{1, *}$\thanks{$^*$ Corresponding author: Fei He (fei.he@coventry.ac.uk). This work is in part supported by the Engineering and Physical Sciences Research Council (EPSRC) under Grant EP/X020193/1.}
\\
$^{1}$Centre for Computational Science and Mathematical Modelling, Coventry University, Coventry, UK\\
$^{2}$Institute for Infocomm Research, A*STAR, Singapore
}

\begin{document}

\maketitle
\thispagestyle{empty}
\pagestyle{empty}

\begin{abstract}
Electroencephalogram (EEG) is a valuable technique to record brain electrical activity through electrodes placed on the scalp. Analyzing EEG signals contributes to the understanding of neurological conditions and developing brain-computer interface. Graph Signal Processing (GSP) has emerged as a promising method for EEG spatial-temporal analysis, by further considering the topological relationships between electrodes. However, existing GSP studies lack interpretability of electrode importance and the credibility of prediction confidence. This work proposes an EEG Graph Mutual Attention Convolutional Network (EEG-GMACN), by introducing an ’Inverse Graph Weight Module’ to output interpretable electrode graph weights, enhancing the clinical credibility and interpretability of EEG classification results. Additionally, we incorporate a mutual attention mechanism module into the model to improve its capability to distinguish critical electrodes and introduce credibility calibration to assess the uncertainty of prediction results. This study enhances the transparency and effectiveness of EEG analysis, paving the way for its widespread use in clinical and neuroscience research.
\end{abstract}

\section{Introduction}
Electroencephalogram (EEG) is a physiological technique used to record brain electrical activity \cite{Wang2024}. By placing electrodes on the scalp, neuronal electrical activity is measured and recorded. Spectrum analysis and spatial pattern recognition of EEG signals may contribute to our understanding of the physiological basis of conditions, such as epilepsy, cognitive disorders, and other neurological disorders. The application of Graph Signal Processing (GSP) in the analysis of EEG \cite{Klepl2024} is gaining widespread attention and demonstrating significant potential. Traditional EEG analysis has primarily focused on temporal and frequency domain features, whereas GSP, by considering the topological relationships between electrodes and introducing graph structures, provides a novel analytical perspective for modeling brain networks \cite{Miri2024}. The advantage of this approach lies in its ability to better capture the topological structure, information propagation, and dynamic regulation within brain networks. The development of GSP in this emerging field holds promise for innovative research and diagnostic tools in neuroscience and clinical neurology.

The current state of GSP in EEG analysis suggests that by integrating graph structure information, we can interpret EEG signals more comprehensively and accurately, enhancing our understanding of brain function and the physiological basis of neurological disorders. However, for existing EEG-based GSP studies, a notable deficiency lies in the lack of in-depth discussions regarding the interpretability \cite{Hanif2023} of electrode weights and the calibration \cite{Ao2023} of model confidence: (1) The interpretability of electrode weights is crucial for revealing the model's reasoning biases and key feature selection; (2) subsequently, an analysis of the interpretability of electrode weights can provide more intuitive and specific auxiliary references for clinical disease diagnosis; (3) and finally, the evaluation of model confidence calibration is essential for discerning the uncertainty of model predictions. Models with inadequate confidence calibration may be prone to misclassify outliers or ambiguous samples.

In this work, we propose an Interpretable EEG Graph Mutual Attention Convolutional Network (EEG-GMACN). This network enhances the interpretability and clinical relevance of EEG inference results based on GSP. It encodes EEG signals into a graph signal format based on the spatial positions of electrodes, and then train a GSP classification model incorporating Graph Convolutional Networks (GCN) \cite{Zhou2020} and attention mechanisms \cite{GAT}. Additionally, this network also includes an interpretability module that elucidates the electrodes weights to by the GSP model during inference by leveraging inverses of weighted graphs and gradient-weighted outputs, providing interpretable electrode graph weights. In order to evaluate the uncertainty of model classification, we introduce confidence calibration into the evaluation metrics. 

\begin{figure*}[!thb]
	\centerline{\includegraphics[width=0.93\linewidth]{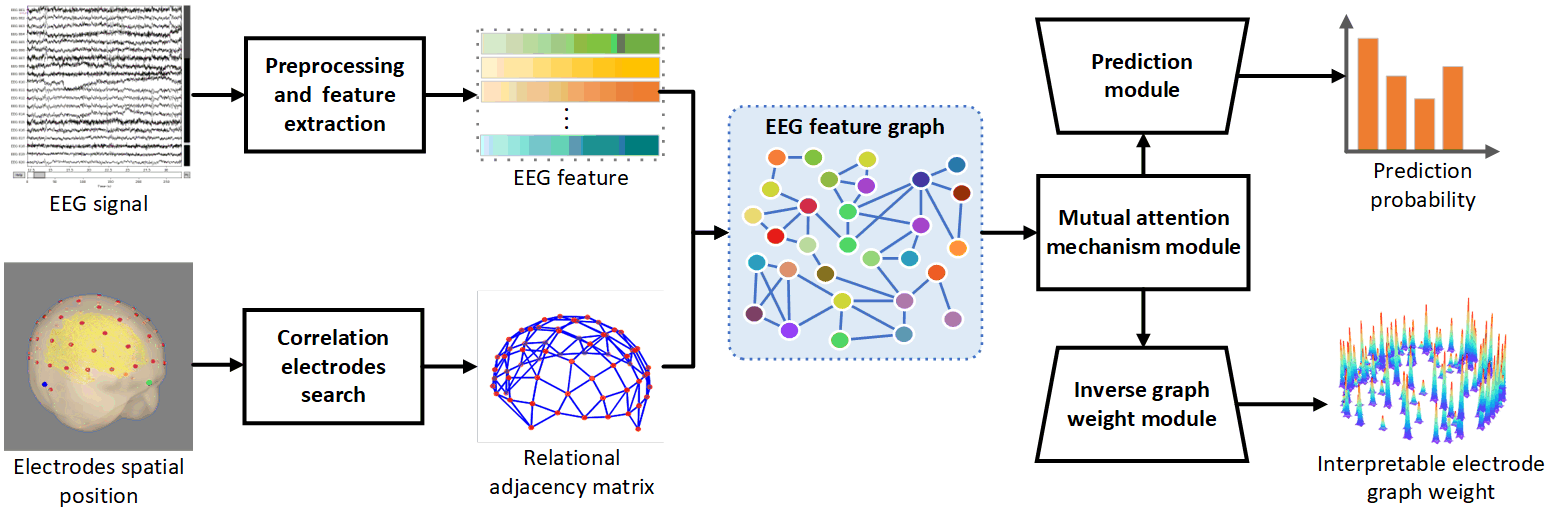}}
	\caption{Overview of GSP Method with Interpretability Module for Revealing Electrode Weights in EEG Recognition Model.}
	\label{fig1}
\end{figure*}

\section{Method}
The overview of the proposed EEG-GMACN  is illustrated in Fig.\ref{fig1}. Firstly, EEG signal undergo preprocessing for standardization and feature extraction through, e.g. wavelet transformation, to capture the time-frequency domain EEG feature. Subsequently, we model the relative spatial differences based on the electrodes' position as an relational adjacency matrix, forming an EEG feature graph alongside the extracted EEG feature. Finally, we employ a mutual attention mechanism module to extract deep representations of the graph, predicting category probabilities for downstream tasks. Particularly, we design a reverse graph weight module that calculates electrode weights by backpropagating gradients and weighting the inverse of the graph. The following content will introduce the specific details of the modules in EEG-GMACN.

\subsection{Preprocessing and Feature Extraction}

In the preprocessing stage, EEG signal samples in this study undergo bandpass filtering between 0.05 and 200 Hz, followed by digitization at 16 bits (0.1 uV precision) at a rate of 1000 Hz. Subsequently, the signals are resampled to 100 Hz to regularize the signal frequency. Next, independent component analysis (ICA) and wavelet transformation (WT) are applied to the EEG signals from each electrode to extract time-frequency features for each channel. This preprocessing workflow aims to reduce noise interference, enhance the time-frequency characteristics of the signals, and provide clearer and more reliable EEG signals for subsequent analysis. Finally, according to the specific event or stimulus marker corresponding to the EEG signal, the corresponding epoch fragments are extracted as EEG feature $V$.

\subsection{Correlation Electrodes Search}
The correlation electrodes search aims to treat each electrode as a node in GSP and establish a relational adjacency matrix based on the relative displacement between electrodes, as shown in Fig.\ref{fig1}. Firstly, standard electrode positions are exported from the international 10-05 system based on the electrode channel names in the EEG signals. Then, through the Graph-based nearest neighbor search method, it searches for neighboring electrodes for each electrode, creating edges in the graph, and thus obtaining a complete relational adjacency matrix. Specifically, this paper discusses two strategies for searching neighbors of electrodes: (1) distance threshold search, and (2) top-k nearest neighbor search.

For the distance threshold search method, distances between all pairs of electrodes are first calculated. Then, a distance threshold $t$ is set, and an edge is considered between electrodes whose distance is less than the threshold $t$. This process can be summarized by the following formula:
\begin{equation}
\check{E}_{i,j} = \frac{1}{1 + \frac{d(e_i, e_j)}{t}},
\end{equation}
here, \(\check{E}_{i,j}\) represents the weight of the edge between electrodes \(e_i\) and \(e_j\). The value of \(d(e_i, e_j)\) corresponds to the euclidean distance between electrodes \(e_i\) and \(e_j\), and \(t\) is the distance threshold. Additionally, the top-k nearest neighbor search is based on identifying the top-k nearest neighboring electrodes to each electrode, which are considered as having edges. This process can be summarized by the following formula:
\begin{equation}
\hat{E}_{i,j} = \frac{1} {1+\frac{\text{min}(\text{rank}(d(e_i, e_j)), k)}{k}},
\end{equation}
where \(\hat{E}_{i,j}\) represents the weight of the edge between electrodes \(e_i\) and \(e_j\), \(\text{rank}(d(e_i, e_j))\) indicates the rank of the euclidean distance between electrodes \(e_i\) and \(e_j\) among all electrode distances, and \(k\) is the specified top-k nearest neighbor count. This formula normalizes the edge weight as the ratio of the minimum between the distance rank and \(k\), ensuring the weight falls within the range of \(0\) and \(1\). Based on these two correlation electrode search strategies, we obtain the adjacency matrix \(E\), which, together with the EEG features \(V\) obtained in the previous section, forms the EEG feature graph \(G = (V, E)\). The performance differences between the two search strategies will be discussed in the experimental section.

\subsection{Mutual Attention Mechanism Module}
The Mutual Attention Mechanism Module (MAMM) aims to enhance the ability of EEG-GMACN to distinguish critical electrode weights by computing attention scores for each pair of electrodes. Specifically, it contains several pairs GCN layer and mutual attention layer. The GCN layer $F_{gcn}$ is used for implicit feature aggregation learning between adjacent nodes. The mutual attention layer \(F_{ma}\) builds upon the output of the GCN layer \(F_{gcn}\). The process of GCN can be expressed as:
\begin{equation}
    F_{gcn}\{H^{(l)}\} =  ReLU\left(\hat{D}^{-\frac{1}{2}}\hat{E}\hat{D}^{-\frac{1}{2}}H^{(l)}W_{gcn}^{(l)}\right),
\end{equation}
where \( H^{(l)} \) represents the node features at layer \( l \), and \( \tilde{E} = E + I \) is the symmetric adjacency matrix formed by adding self-connections to the original adjacency matrix \( E \), \( \tilde{D} \) is the diagonal degree matrix of \( \tilde{E} \). Moreover, \( W_{gcn}^{(l)} \) is the learnable weight matrix at layer \( l \). On this basis, if we denote the output of the GCN at layer \( l \) as \( O^l \), the output of mutual attention module layer \(F_{ma}\) within the same layer can be expressed as:
\begin{equation}
F_{ma}\{O^l\} = \text{softmax}(O^l \cdot W_q \cdot (O^l \cdot W_k)^T) \cdot (O^l \cdot W_v),
\end{equation}
where \(W_q\), \(W_k\), and \(W_v\) are learnable weight matrices, and \(\text{softmax}\) is applied along the appropriate axis to obtain normalized attention coefficients. Hence, the complete process of inputting \(H^l\) into the MAMM to output \(H^{l+1}\) can be expressed as:
\begin{equation}
		H^{(l+1)} = F_{gcn}\{H^{(l)}\} \odot F_{ma}\{F_{gcn}\{H^{(l)}\}\},
\end{equation}
where \(\odot\) denotes element-wise multiplication. The output at the \((l+1)\)-th layer, denoted as \(H^{(l+1)}\), is obtained by applying the GCN layer \(F_{gcn}\) to the node feature matrix at the \(l\)-th layer \(H^{(l)}\). The element-wise multiplication \(\odot\) is used to modulate the output of the GCN layer with the attention weights obtained from the mutual attention layer \(F_{ma}\). This helps to learn the importance of each electrode with respect to the entire graph. The EEG feature graph, after undergoing the mamm extraction, have deep implicit features fed into two branches: (1) a prediction module, and, (2) an inverse graph weight module. In the prediction module, the features extracted from MAMM are concatenated into a feature vector, which is then sequentially inputted into FC layers and Softmax to obtain the predicted probabilities for each class. During the training phase, the model's loss function combines cross-entropy loss and focal loss \cite{Lin2017}.

\subsection{Inverse Graph Weight Module}
The graph weight inverse module analyzes the influence weights of each electrode on the predicted results by calculating the graph feature gradient and the graph weight inverse output interpretable electrode graph weights. For each class probability output \(P=\{p_1, p_2, ..., p_N\}\) of the model, $N$ indicates the total number of categories. The class \(c\) with the highest output probability \(P_c\) is selected as the target class. Subsequently, based on Grad-CAM \cite{Selvaraju2019}, the graph feature gradients $\text{GFG}(v)$ with respect to the first GCN layer for the target class are obtained: 
\begin{equation}
\text{GFG}(v_i) = \frac{\partial P(c | F_{gcn}^{(1)})}{\partial H^{(1)}} \cdot H^{(1)}(v_i),
\end{equation}
where \(H^{(1)}(v_i)\) denotes feature of node \(v_i\) in the first GCN layer. Then, we represent the inverse weight of the graph-weighted inverses $\text{GWI}(v)$ by separately calculating the gradient weight of the first mutual attention layer:
\begin{equation}
\text{GWI}(v_i) = \frac{\partial P(c | F_{ma}^{(1)})}{\partial A^{(1)}} \cdot A^{(1)}(v_i),
\end{equation}
where \(A^{(1)}(v_i)\) denotes node feature of \(v_i\) in the first mutual attention layer. Finally, we combine the graph feature gradients $\text{GFG}(v)$, graph-weighted inverses $\text{GWI}(v)$, and the adjacency matrix $E$ in a weighted manner to obtain the ultimate interpretable electrode graph weights $\text{IEGW}(v)$:
\begin{equation}
	\text{IEGW}(v_i) = Normal(\text{GFG}(v) \odot \text{GWI}(v) \odot E),
\end{equation}
We utilized Min-Max Normalization $Normal()$ to map the original weights to a range between 0 and 1, reducing the impact of outliers on the normalization effect.

\section{Experiments}
\subsection{Datasets}
This paper conducted experiments on the BCI III dataset \cite{BCI-III} to validate the effectiveness of the proposed method. EEG signals were collected from two subjects while they were fixating on a letter matrix (170 training samples and 200 testing samples). The task in this dataset required the model to predict the letter category (26 kinds of characters) that the subject was fixating on based on the EEG signals. The dataset comprises 64 electrodes.
\subsection{Baseline}
We selected four state-of-the-art EEG classification models (EEGNet\cite{EEGNet}, Im-EEGNet\cite{Improved-EEGNet}, EEG-GCNN\cite{EEG-GCNN} and EEG-GNN\cite{EEG-GNN}) as comparative baselines. Among them, EEG-GCNN\cite{EEG-GCNN} and EEG-GNN\cite{EEG-GNN} are EEG classification models based on GSP, they employ GNN or GCN models for EEG signal classification. The difference is that EEG-GMACN introduces the Inverse Graph Weight Module to further explore the interpretability of electrode weights in GSP models for EEG classification tasks. 
\vspace{-1.0em}
\begin{table}[h]
	\caption{Comparison of Model Performance Using Different Correlation Electrodes Search Strategies}\label{tab1}
	\centering
	\begin{tabular}{c|ccc|cc}
		\hline
		$t$ & $S_{Acc} \uparrow$ & $S_{Pre} \uparrow$ & $S_{Rec} \uparrow$ & $S_{F1} \uparrow$ & $S_{ECE}\downarrow$\\
		\hline
		10 & 0.8123 & 0.8054 & 0.7985 & 0.8102 & 0.1342 \\
		\textbf{20} & \textbf{0.8257} & \textbf{0.8163} & \textbf{0.8423} & \textbf{0.8154} & \textbf{0.0764}\\
		30 & 0.7892 & 0.7943 & 0.7685 & 0.7921 & 0.0987 \\ 
		\hline
		$k$ & $S_{Acc}$ & $S_{Pre}$ & $S_{Rec}$ & $S_{F1}$ & $S_{ECE}$ \\
		\hline
		3 & 0.7645 & 0.7871 & 0.7452 & 0.7689 & 0.1538 \\
		5 & 0.7912 & 0.8043 & 0.7889 & 0.7981 & 0.1265 \\
		7 & 0.7783 & 0.7956 & 0.7812 & 0.7891 & 0.1421 \\
		\hline
	\end{tabular}
	\label{tab1}
\end{table}
\vspace{-1.0em}
\subsection{Evaluation Metrics}
The model's performance is evaluated using accuracy ($S_{Acc}$), precision ($S_{Pre}$), recall ($S_{Rec}$), F1 score ($S_{F1}$), and Expected Calibration Error (ECE) ($S_{ECE}$). ECE serves as a metric for assessing probability calibration performance. ECE quantifies the disparity between the model's predicted uncertainty and the actual observed accuracy, providing a reliable calibration measure. Additionally, Flops ($F(M)$) and Params ($P(M)$) are used to evaluate the computational complexity of the model in structural ablation experiments.

\subsection{Ablation Study}
As shown in Table \ref{tab1}, we present comparative experimental results for two different search strategies in BCI III test set. Their classification modules are uniformly designed as three-layer GCN without incorporating mutual attention layers \(F_{ma}\). For the distance threshold-based method, we assess the performance when the distance threshold $t$ is set to 10, 20, and 30 pixel. Regarding the top-k nearest neighbors method, we evaluate the performance for different values of $k$, specifically 3, 5, and 7. The experimental results indicate that the optimal performance is achieved when using the distance threshold method with a threshold \( t \) set to 20. The $F_1$ Score and ECE reach 0.8154 and 0.0764, respectively. 
\vspace{-0.5em}
\begin{table}[h]
	\caption{Comparison of Model Performance Using Mutual Attention Layers $F_{ma}$}\label{tab2}
	\centering
	\begin{tabular}{p{0.4cm}|p{0.7cm}p{0.7cm}p{0.7cm}p{0.7cm}p{0.9cm}|p{0.7cm}p{0.8cm}}
		\hline
		$F_{ma}$ & $S_{Acc} \uparrow$ & $S_{Pre} \uparrow$ & $S_{Rec} \uparrow$ & $S_{F1} \uparrow$ & $S_{ECE}\downarrow$ & $F(M)\downarrow$ & $P(M)\downarrow$\\
		\hline
		& 0.8257 & 0.8163 & 0.8423 & 0.8154 & 0.0764 & \textbf{2.67} & \textbf{0.99}\\
		\textbf{\checkmark} & \textbf{0.8352} & \textbf{0.8276} & \textbf{0.8503} & \textbf{0.8371} & \textbf{0.0658} & 4.46 & 1.45\\
		\hline
	\end{tabular}
	\label{tab2}
\end{table}

\vspace{-0.7em}
\begin{figure}
    \centerline{\includegraphics[width=0.80\linewidth]{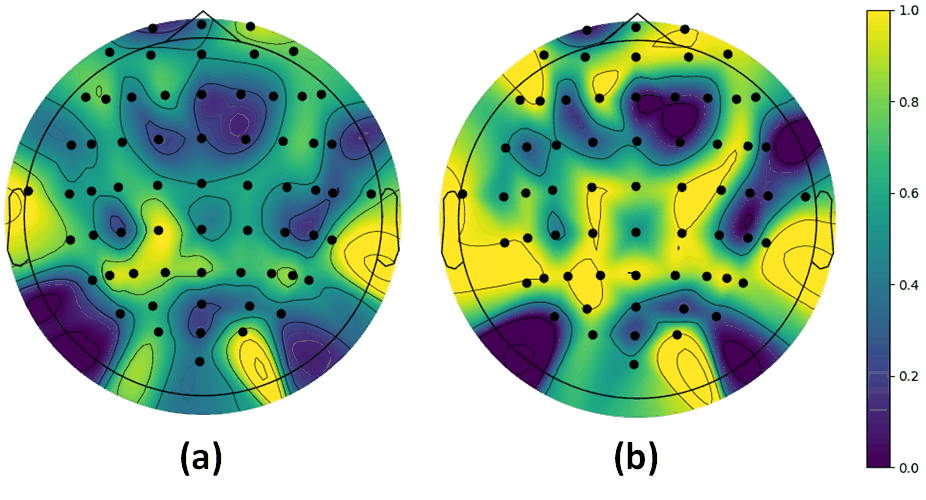}}
	\caption{Visual Comparison of Electrode Graph Weight Heatmaps Exported using the IEGW Module on the BCI III Test Set. (a) w/o mutual attention layers in the model, and (b) w/ mutual attention layers in the model.}
	\label{fig2}
\end{figure}

\vspace{-1.0em}

\begin{table}[h]
	\caption{State-of-the-Art (SOTA) Model Comparison on BCI III Test Set}\label{tab3}
	\centering
	\begin{tabular}{c|ccccc}
		\hline
		Model & $S_{Acc} \uparrow$ & $S_{Pre} \uparrow$ & $S_{Rec} \uparrow$ & $S_{F1} \uparrow$ & $S_{ECE}\downarrow$\\
		\hline
		EEGNet\cite{EEGNet} & 0.7982 & 0.7914 & 0.7837 & 0.7956 & 0.1503 \\
		Im-EEGNet \cite{Improved-EEGNet} & 0.8101 & 0.8026 & 0.7949 & 0.8073 & 0.1367 \\
		EEG-GCNN \cite{EEG-GCNN} & 0.8165 & 0.8087 & 0.8010 & 0.8136 & 0.1242 \\
		EEG-GNN \cite{EEG-GNN} & 0.8257 & 0.8178 & 0.8101 & 0.8232 & 0.1098 \\
		\textbf{EEG-GMACN} & \textbf{0.8352} & \textbf{0.8276} & \textbf{0.8503} & \textbf{0.8371} & \textbf{0.0658}\\
		\hline
	\end{tabular}
	\label{tab3}
\end{table}

Furthermore, we uniformly employed the distance threshold-based search method with \( t \) set to 20 to compare the performance differences of the model after incorporating the mutual attention layers $F_{ma}$. The experimental results are presented in Table \ref{tab2}. The experimental results indicate a significant improvement in the model's performance metrics upon introducing $F_{ma}$. The F1 and ECE achieved notable enhancements, reaching 0.8371 and 0.0658, respectively. Despite the added computational load of 1.79M FLOPs and 0.46M Params, the operational efficiency remains sufficient for real-time clinical detection needs. Fig.\ref{fig2} illustrates a visual comparison of electrode graph weight heatmaps exported using the IEGW module. It is obvious that the addition of $F_{ma}$makes the model more discriminating between features from different electrode channels, which can help the model pay more attention to features from key electrodes. After validation, it was observed that the model achieves optimal performance when the number of layers in the MAMM is set to 3, and a distance threshold-based method with a threshold of $t=20$ is employed. This configuration will be adopted as the final implementation of EEG-GMACN.

\subsection{Comparison Experiment}
As illustrated in Table \ref{tab3}, we conducted an extensive evaluation of our proposed EEG-GMACN against state-of-the-art models on the BCI III test set. The results showed that EEG-GMACN consistently outperformed existing models, showcasing its robustness and superiority in EEG signal classification tasks. Especially when compared to similar GSP methods EEG-GCNN\cite{EEG-GCNN} and EEG-GNN\cite{EEG-GNN}, our approach demonstrated improvements of 0.0235 and 0.0139 in F1 score and reductions of 0.0584 and 0.0440 in ECE.

\section{Conclusion}
In summary, our proposed EEG-GMACN combines mutual attention with an interpretability module, aiming to reveal electrode weights in EEG classification models. Overall, our work contributes to enhancing the transparency and effectiveness of EEG model applications, paving the way for widespread use in clinical and neuroscience research. Future work may explore the feasibility of lightweight interpretable EEG graph single processing networks.

\bibliographystyle{IEEEtran}
\bibliography{ref}

\end{document}